\documentclass[prl,
twocolumn,
showpacs,
]{revtex4}
\usepackage{graphicx}

\begin{document}

\title{Different routes to charge disproportionation in perovskites-type Fe oxides}

\author{J.~Matsuno}
\altaffiliation{Present address: Correlated Electron Research Center
  (CERC), National Institute of Advanced Industrial Science and
  Technology (AIST), Tsukuba 305-8562, Japan}
\affiliation{Department of Physics and Department of Complexity Science and Engineering, University of Tokyo,
  Bunkyo-ku, Tokyo 113-0033, Japan}

\author{T.~Mizokawa}
\affiliation{Department of Physics and Department of Complexity Science and Engineering, University of Tokyo,
  Bunkyo-ku, Tokyo 113-0033, Japan}

\author{A.~Fujimori}
\affiliation{Department of Physics and Department of Complexity Science and Engineering, University of Tokyo,
  Bunkyo-ku, Tokyo 113-0033, Japan}

\author{Y.~Takeda}
\affiliation{Department of Chemistry, Mie University,
Tsu 514-8507, Japan}

\author{S.~Kawasaki}
\affiliation{Institute of Chemical Research, Kyoto University,
Uji, Kyoto 611-0011, Japan}

\author{M.~Takano}
\affiliation{Institute of Chemical Research, Kyoto University,
Uji, Kyoto 611-0011, Japan}

\date{\today}

\begin{abstract}
  Iron perovskites CaFeO$_3$ and La$_{0.33}$Sr$_{0.67}$FeO$_3$ show
  charge disproportionation, resulting in charge-ordered states with
  Fe$^{3+}$:Fe$^{5+}$ $=1:1$ and $=2:1$, respectively. We have made
  photoemission and unrestricted Hartree-Fock band-structure
  calculation of CaFeO$_3$ and compared it with
  La$_{0.33}$Sr$_{0.67}$FeO$_3$. With decreasing temperature, a
  gradual decrease of the spectral weight near the Fermi level
  occurred in CaFeO$_3$ as in La$_{0.33}$Sr$_{0.67}$FeO$_3$ although
  lattice distortion occurs only in CaFeO$_3$. Hartree-Fock
  calculations have indicated that both the breathing and tilting
  distortions are necessary to induce the charge disproportionation in
  CaFeO$_3$, while no lattice distortion is necessary for the charge
  disproportionation in La$_{0.33}$Sr$_{0.67}$FeO$_3$.
\end{abstract}

\maketitle 

Recently charge ordering phenomena in transition-metal oxides have
been extensively studied, particularly in relation to charge stripes
in the high-$T_C$ cuprates~\cite{cuprates} and giant magnetoresistance
in the manganites~\cite{manganites}.  Charge ordering in highly
covalent transition-metal oxides containing $d^4$ (Fe$^{4+}$,
Mn$^{3+}$) or $d^7$ (Ni$^{3+}$) ions often exhibit so-called charge
disproportionation, in which a charge state is thought to be separated
into two different charge states as $2d^n \to d^{n-1} + d^{n+1}$. A
clear fingerprint of charge disproportionation is the breathing-type
distortion of metal-oxygen octahedra since the different charge states
of the transition-metal ion take different ionic radii. Indeed, a
breathing-type lattice distortion has been found for
YNi$^{3+}$O$_3$~\cite{YNiO3CD}, NdNi$^{3+}$O$_3$~\cite{NdNiO3CD} and
CaFe$^{4+}$O$_3$ by neutron diffraction studies~\cite{Woodward,Takeda}
and a thermally fluctuating charge disproportionated state has been
postulated for LaMnO$_3$ at high temperatures~\cite{LaMnO3}. However,
charge disproportionation in La$_{1-x}$Sr$_x$FeO$_3$ with $x\simeq$
0.7, which occurs as a first-order phase transition from the
paramagnetic average-valence state (Fe$^{\sim3.7+}$) above 200 K to
the antiferromagnetic charge-disproportionated state
(Fe$^{3+}$:Fe$^{5+}$ $=2:1$) below that temperature~\cite{Takano}, is
not accompanied by an appreciable lattice distortion~\cite{Battle}
although an electron diffraction study has shown extra spots~\cite{Li}
and two charge states of Fe with different hyperfine fields have been
detected by M\"ossbauer spectroscopy~\cite{Takano}.  The neutron
diffraction study of La$_{0.3}$Sr$_{0.7}$FeO$_3$~\cite{Battle} has
revealed a spin-density wave (SDW) of six-fold periodicity along the
$\langle$111$\rangle$ direction with two inequivalent spin states of
Fe, suggesting a charge-density wave (CDW) of three-fold periodicity
along the same direction.

Recent photoemission studies have revealed that the electronic
structure of Fe$^{4+}$ oxides is rather unique: the charge-transfer
energy $\Delta$, the energy required to transfer an electron from the
oxygen $p$ to the Fe $3d$ level, is extremely negative ($\sim$ $-$3 eV
including Hund's coupling energy)~\cite{SFO_Bocquet}, and the ground
state of the formal Fe$^{4+}$ (``$d^4$'') state is in fact dominated
by the $d^5\underline{L}$ configuration, where $\underline{L}$ denotes
a hole in the oxygen 2$p$ band.  The charge disproportionation is
therefore more correctly described as $2d^5\underline{L} \to
d^5\underline{L}^2 + d^5$ rather than $2d^4 \to d^3 + d^5$.  To
understand the interesting physical properties of Fe$^{4+}$ oxides
such as the helical antiferromagnetic metallic state in
SrFeO$_3$~\cite{SrFeO3}, the Co-substitution induced ferromagnetism in
SrFe$_{1-x}$Co$_x$O$_3$~\cite{Kawasaki, Abbate}, and magnetic and
electric phase transitions under high pressure~\cite{Takano2,
  Kawakami, Rozenberg}, therefore, one has to take into account the
negative $\Delta$, namely, the oxygen-hole character of charge
carriers. Especially, the $d^5\underline{L}$ ground state naturally
explains the fact that SrFeO$_3$ shows no charge disproportionation
yet no Jahn-Teller effect since the $d^4$ ion necessarily undergoes a
Jahn-Teller distortion.

In a recent photoemission and unrestricted Hartree-Fock band-structure
calculation study of La$_{1-x}$Sr$_x$FeO$_3$~\cite{Matsuno}, we have
shown that the charge disproportionation is purely electronically
driven and that the ordering of oxygen holes plays an important role
in the charge disproportionated state of
La$_{0.33}$Sr$_{0.67}$FeO$_3$.  In this Letter, we present a
photoemission study of CaFeO$_3$, which unlike
La$_{0.33}$Sr$_{0.67}$FeO$_3$ shows a breathing-type lattice
distortion in the charge disproportionated state. Using photoemission
we have studied how the charge disproportionation influences the
electronic structure near the Fermi level ($E_F$) in comparison with
La$_{0.33}$Sr$_{0.67}$FeO$_3$. We have also performed unrestricted
Hartree-Fock band-structure calculations on CaFeO$_3$ taking into
account the realistic lattice distortion. Based on those results,
different driving mechanisms are proposed for the charge
disproportionation in the seemingly very similar systems CaFeO$_3$ and
La$_{0.33}$Sr$_{0.67}$FeO$_3$.

\begin{figure}[ht]
 \begin{center}
  \includegraphics[scale=0.6]{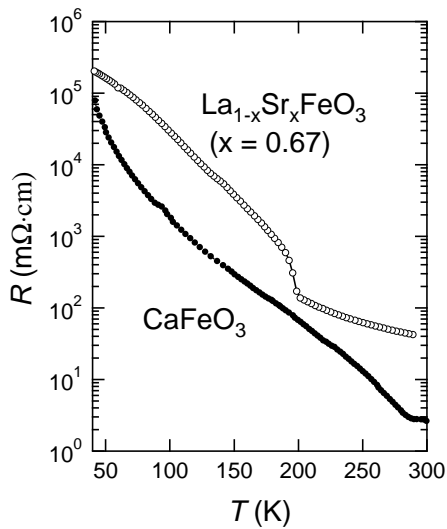}
  \caption{Electrical resistivities of CaFeO$_3$ and
    La$_{0.33}$Sr$_{0.67}$FeO$_3$.}
  \label{fig:0}
 \end{center}
\end{figure}

A polycrystalline sample of CaFeO$_3$ was prepared by a solid state
reaction and a subsequent treatment under high-pressure
oxygen~\cite{Kawasaki}. The electrical resistivity of the present
CaFeO$_3$ sample is shown in Fig.~\ref{fig:0} and is compared with
that of La$_{1-x}$Sr$_{x}$FeO$_3$ ($x$ = 0.67)~\cite{Matsuno}.  It
shows a gradual increase below the anomaly at 290 K due the gradual
charge disproportionation 2Fe$^{4+}$ $\rightarrow$ Fe$^{5+}$ $+$
Fe$^{3+}$ while no anomaly due to the antiferromagnetic ordering is
seen at the N\'eel temperature $T_N=$ 115 K.

Ultraviolet photoemission (UPS) measurements were made using the He
{\footnotesize I} resonance line ($h\nu=$ 21.2 eV). The He
{\footnotesize I} spectra have been corrected for the He
{\footnotesize I}$^{*}$ satellite. In order to calibrate binding
energies and to estimate the instrumental resolution, gold was
evaporated on the sample surface after each series of measurements.
The energy resolution was 48 meV.  The samples were repeatedly scraped
{\it in situ} with a diamond file.  We have adopted the spectra taken
within 40 minutes after scraping and the reproducibility of the
spectra was confirmed by repeated measurements.

\begin{figure}[ht]
 \begin{center}
  \includegraphics[scale=0.5]{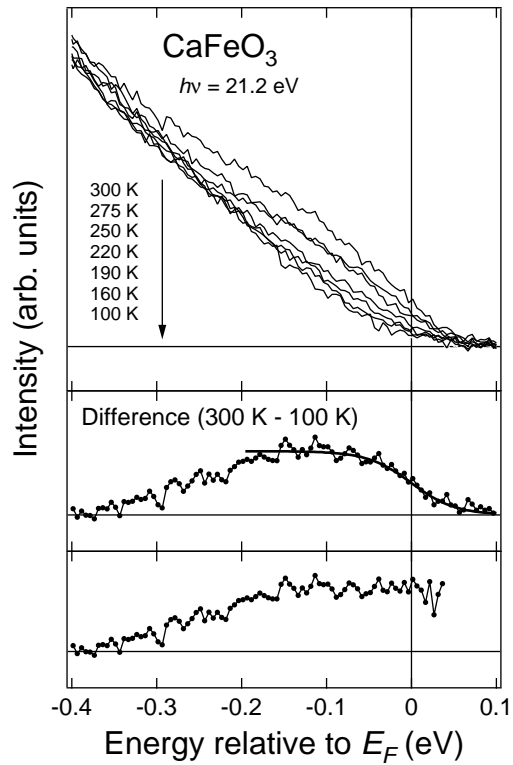}
  \caption{Photoemission spectra of CaFeO$_3$. Top: Spectra taken at various
    temperatures. Middle: Difference spectrum between 300 K and 100 K.
    The solid curve is the Fermi-Dirac distribution function at 300 K
    convoluted with the instrumental resolution. Bottom: The same
    difference spectrum divided by the Fermi-Dirac distribution
    function.}
  \label{fig:1}
 \end{center}
\end{figure}
\begin{figure}[ht]
 \begin{center}
  \includegraphics[scale=0.5]{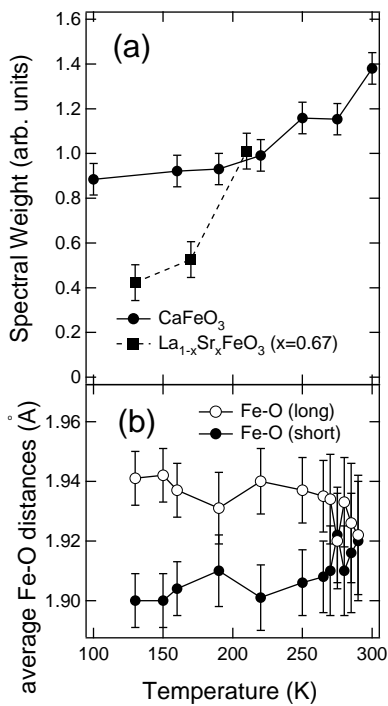}
  \caption{(a) Spectral weight integrated from $-$0.20 to $+$0.05 eV for 
    CaFeO$_3$ and La$_{0.33}$Sr$_{0.67}$FeO$_3$~\cite{Matsuno} as
    functions of temperature. (b) Average Fe-O bond lengths observed
    by neutron diffraction~\cite{Takeda}.}
  \label{fig:2}
 \end{center}
\end{figure}
Figure~\ref{fig:1} shows the temperature dependence of the He
{\footnotesize I} spectra of CaFeO$_3$ near $E_{F}$. The spectra have
been normalized to the integrated intensity of the entire valence
band.  A clear change in the intensity from the Fermi level to 0.4 eV
below it has been observed as in the case of
La$_{0.33}$Sr$_{0.67}$FeO$_3$. One can see that the spectral weight
near $E_F$ show a gradual decrease of the spectral weight with
decreasing temperature.  The most prominent temperature dependence
occurs just below 300 K, followed by more gradual changes below
$\sim$270 K.  In order to illustrate this, we have plotted the
spectral weight integrated from $-$0.20 to $+$0.05 eV as a function of
temperature in Fig.~\ref{fig:2}(a), where the previous result on
La$_{0.33}$Sr$_{0.67}$FeO$_3$ is also plotted~\cite{Matsuno}. One can
see a more gradual decrease as a function of temperature in CaFeO$_3$
than in La$_{0.33}$Sr$_{0.67}$FeO$_3$ corresponding to the more
gradual change in the electrical resistivity below the charge
disproportionation temperature, although the total spectral change in
the wide temperature range is similar between the two systems. We also
note that this temperature dependence of the spectral weight is
qualitatively similar to the temperature dependence of the Fe-O bond
length~\cite{Takeda} as shown in Fig.~\ref{fig:2}(b), suggesting that
the lattice deformation is related to the change of the photoemission
spectra in CaFeO$_3$.  It is remarkable that, although
La$_{0.33}$Sr$_{0.67}$FeO$_3$ shows negligibly small structural
changes across the transition~\cite{Battle}, it shows spectral changes
as strong as CaFeO$_3$. These observations imply different mechanisms
for the charge disproportionation between the two systems.

In Fig.~\ref{fig:1}, the difference spectrum between 300 K and 100 K
is compared with the Fermi-Dirac distribution function $f_D(\epsilon)$
at $T$ = 300 K. The difference spectrum seems to agree rather well
with $f_D(\epsilon)$ near $E_F$. This is different from
La$_{0.33}$Sr$_{0.67}$FeO$_3$, which shows a pseudo-gap-like behavior
at $E_F$. In order to clarify this, the difference spectrum divided by
the Fermi-Dirac distribution function is also shown in the bottom
panel. The obtained spectral DOS is nearly flat around the $E_F$, in
contrast to the corresponding spectrum for
La$_{0.33}$Sr$_{0.67}$FeO$_3$, where a pseudo-gap-like DOS has been
observed~\cite{Matsuno}. This is consistent with the transport
properties, which show more conducting and metallic behavior in
CaFeO$_3$ than in La$_{0.33}$Sr$_{0.67}$FeO$_3$ above the transition
temperature.

In order to study the driving force for the phase transition and the
origin of the band gap below the transition temperature in CaFeO$_3$,
we have carried out unrestricted Hartree-Fock calculations for the
multi-band $d$-$p$ lattice model, in which the full degeneracy of the
Fe 3$d$ and oxygen 2$p$ orbitals are taken into
account~\cite{Mizokawa1}. Parameters in the model are the
charge-transfer energy $\Delta$, the multiplet averaged $d$-$d$
Coulomb interaction $U$ and Slater-Koster parameters
$(pd\sigma),(pd\pi),(pp\sigma)$ and $(pp\pi)$, which represent
transfer integrals between the transition metal 3$d$ and oxygen 2$p$
orbitals. The values of $\Delta, U$ and $(pd\sigma)$ were chosen to be
0, 6 and $-1.8$ eV, respectively. The ratio $(pd\sigma)/(pd\pi)$ was
fixed at $-2.16$ and $(pp\sigma)$ and $(pp\pi)$ at $-0.60$ and 0.15
eV, respectively~\cite{Mattheiss}.

\begin{figure}[ht]
 \begin{center}
  \includegraphics[scale=0.5]{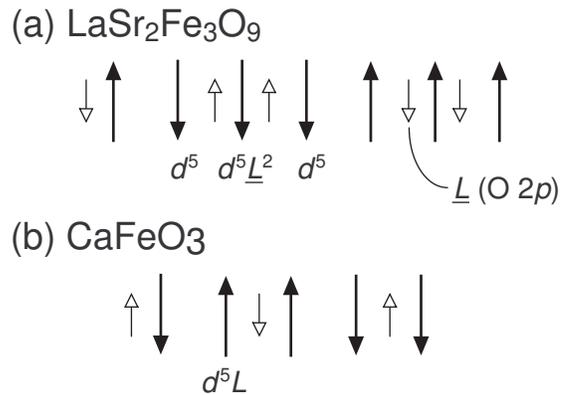}
  \caption{Schematic descriptions of the spin and charge
    configurations in LaSr$_2$Fe$_3$O$_9$
    (La$_{0.33}$Sr$_{0.67}$FeO$_3$) (a) and CaFeO$_3$ (b). Black and
    white arrows represent the spins at the Fe sites and those of the
    oxygen holes, respectively.}
  \label{fig:Fe_CO}
 \end{center}
\end{figure}

First, in analogy with LaSr$_2$Fe$_3$O$_9$ (a supercell model for
La$_{0.33}$Sr$_{0.67}$FeO$_3$)~\cite{Matsuno}, where charge ordering
with Fe$^{3+}$ : Fe$^{5+}$ $=2:1$ causes an SDW with six-fold
periodicity along the $\langle$111$\rangle$ direction
[Fig.~\ref{fig:Fe_CO}(a)], we assumed an SDW with four-fold
periodicity, namely $\uparrow\uparrow\downarrow\downarrow$, for
CaFeO$_3$, where ordering with Fe$^{3+}$ : Fe$^{5+}$ $= 1:1$ occurs in
the same direction. This assumption is compatible with the alternating
charge order of Fe$^{3+}$ and Fe$^{5+}$ in CaFeO$_3$. Under this
assumption we obtained an insulating solution but this was not a
disproportionated one. The reason why we could not explain both
charge-ordered and insulating ground state within the
$\uparrow\uparrow\downarrow\downarrow$ model is schematically shown in
Fig.~\ref{fig:Fe_CO}(b). The tendency that holes enter oxygen orbitals
between the iron sites of the parallel spins also holds in this case,
but all the iron sites become equivalent in spite of the hole ordering
in this geometry, as can be seen from Fig.~\ref{fig:Fe_CO}(b). Thus
the origin of the disproportionation in CaFeO$_3$ and in
LaSr$_2$Fe$_3$O$_9$ cannot be the same. In other words, the kinetic
exchange interaction derived from the particular magnetic structure as
in the case of La$_{0.33}$Sr$_{0.67}$FeO$_3$ cannot drive the charge
disproportionation in CaFeO$_3$.

Then we introduced the lattice distortion in CaFeO$_3$ explicitly
because in CaFeO$_3$, lattice distortion has been found to be
substantial unlike La$_{0.33}$Sr$_{0.67}$FeO$_3$. CaFeO$_3$ has the
GdFeO$_3$-type tilting distortion as well as the breathing
distortion~\cite{Woodward,Takeda,Morimoto} and the Fe-O-Fe bond angle
is determined to be $\sim$ 158$^{\circ}$. In order to take into
account these distortions, a unit cell has been selected as containing
four FeO$_6$ octahedra. The breathing distortion, which results in two
kinds of Fe-O bond lengths, has been treated by scaling the transfer
integrals with respect to the bond lengths ($d$) following Harrison's rule
~\cite{Harrison} $(pd\sigma) \propto d^{-3.5}$. In order to simulate
the effect of both the breathing and the tilting, we calculated the
magnitude of the band gap as a function of the degree of the breathing
distortion $R =d_{\rm long}/d_{\rm short}$ and the Fe-O-Fe bond angle.
As for the magnetic structure, we assumed the ferromagnetic structures
considering that the spin alignment between neighboring iron sites is
almost ferromagnetic in the screw antiferromagnetic structure with a
long wave vector [= 0.161($111$)2$\pi/a$] observed in
CaFeO$_3$~\cite{Woodward}. The results are shown in
Table~\ref{tab:CFO_GAP}.
\begin{table}[t]
  \begin{tabular}{cccc}\hline \hline
   $\angle$ Fe-O-Fe &$R=1$  &$R=1.022$  &$R=1.043$\\ \hline 
   180$^{\circ}$  & 0  & 0   & 190 \\
   158$^{\circ}$  & 0  & 30   & 300 \\
   \hline \hline\\
  \end{tabular}
  \caption{Band gap in units of meV for various breathing 
   (represented by $R =d_{\rm long}/d_{\rm short}$) and tilting
   (represented by Fe-O-Fe) distortions for
   CaFeO$_3$. $R=1.022$ and Fe-O-Fe $=158^{\circ}$ are 
  the experimental values. }
  \label{tab:CFO_GAP}
\end{table}
We have obtained the charge-disproportionated solution for $R \neq 1$
as expected. The experimental values for CaFeO$_3$ are $R=1.022$ and
$\angle$Fe-O-Fe $=158^{\circ}$ for CaFeO$_3$, however, the breathing
distortion of $R=1.022$ alone or the tilting of $\angle$Fe-O-Fe$ =
158^{\circ}$ alone does not open a gap although Hartree-Fock
calculations tend to overestimate band gaps. In order to open a band
gap (and to lower the total energy of the system), the two types of
lattice distortions should occur simultaneously.  If the bond angle is
180$^{\circ}$ as in the case of SrFeO$_3$, one needs an unreasonably
large breathing distortion to open the gap. In order to confirm the
co-operative nature of the two types of distortions, first-principles
total-energy calculations would be necessary in future work.

The present scenario for the charge disproportionation in CaFeO$_3$
naturally explains why the isoelectronic SrFeO$_3$ remains free from
charge disproportionation: since the ionic radius of Sr$^{2+}$ is
large, no tilting distortion is possible for SrFeO$_3$ and hence the
breathing is also blocked.  This picture may be analogous to the case
of the distorted perovskite BaBiO$_3$, in which freezing of a
breathing phonon mode is well known~\cite{BKBO_review}. Liechtenstein
{\it et al.}~\cite{BKBO} calculated the total energy of BaBiO$_3$ as a
function of tilting and breathing distortions and found that the
instability and the gap opening occurs only when the two kinds of
distortions are combined; in the presence of the tilting distortion,
the nesting instability of the Fermi surface emerges, leading to the
alternating breathing distortion of Bi-O octahedra. If the same
nesting instability is confirmed for CaFeO$_3$ by first-principles
band-structure calculations, then the quite different mechanism of the
charge disproportionation would be established between
La$_{0.33}$Sr$_{0.67}$FeO$_3$, where the ordering of holes at oxygen
sites is purely electronically driven, and CaFeO$_3$, where the
lattice distortion and associated electron-phonon coupling is
important.

In conclusion, we have shown that the driving force for the charge
disproportionation is quite different between the two apparently very
similar compounds La$_{0.33}$Sr$_{0.67}$FeO$_3$ and CaFeO$_3$. The
present finding implies that there exists subtle interplay between
electron-electron interaction, magnetic interaction, and
electron-lattice interaction in realizing charge ordering (or more
generally charge inhomogeneity) and charge fluctuations in
transition-metal oxides. This means that dominant interaction may
change within the same family of compounds, and calls for a critical
re-examination of charge ordering/fluctuation phenomena depending on
the chemical composition (hole concentration) and crystal distortion
even for an apparently similar class of materials.

Discussions with M. Seto and N. Hamada are gratefully acknowledged. 
This work was supported by a Grant-in-Aid for Scientific Research 
(A12304018) from the Ministry of Education, Culture, Sports, Science 
and Technology.


\begin{thebibliography}{0}
\bibitem{cuprates} J.M.~Tranquada, B.J.~Sternlieb, J.D.~Axe,
  Y.~Nakamura, and S.~Uchida, Nature {\bf 375}, 561 (1995).
\bibitem{manganites} See. e.g., {\it Colossal Magnetoresistance,
    Charge Ordering and Related Properties of Manganese Oxides}, ed.
  C.N.R.~Rao and B.~Raveau (World Scientific, Singapore, 1998)
\bibitem{YNiO3CD} J.A.~Alonso, J.L.~Garc\'{i}a-Mu\~{n}oz,
  M.T.~Fern\'{a}ndez-D\'{i}az, M.A.G.~Aranda, M.J.~Mart\'{i}nez-Lope,
  and M.T.~Casais, Phys. Rev. Lett. {\bf 82}, 3871 (1999).
\bibitem{NdNiO3CD}M.~Zaghrioui, A.~Bulou, P.~Lacorre, and P.~Laffez,
  Phys. Rev. B {\bf 64}, 081102(R) (2001).
\bibitem{Woodward} P.M.~Woodward, D.E.~Cox, E.~Moshopoulou,
  A.W.~Sleight, and S.~Morimoto, Phys. Rev. B {\bf 62}, 844 (2000).
\bibitem{Takeda} T.~Takeda, R.~Kanno, Y.~Kawamoto, M.~Takano,
  S.~Kawasaki, T.~Kamiyama, F.~Izumi, Solid State Sci. {\bf 2}, 673
  (2000).
\bibitem{LaMnO3} R.~Raffaelle, H.U.~Anderson, D.M.~Sparlin, and
  P.E.~Parris, Phys.  Rev. B {\bf 43}, 7991 (1991);
  J.A.M.~Van~Roosmalen and E.H.P.~Cordfunke, J. Solid State Chem. {\bf
    110}, 109 (1994)
\bibitem{Takano} M.~Takano, J.~Kawachi, N.~Nakanishi, and Y.~Takeda,
  J. Solid State Chem. {\bf 39}, 75 (1981).
\bibitem{Battle} P.D.~Battle, T.C.~Gibb, and P.~Lightfoot, J. Solid
  State Chem. {\bf 84}, 271 (1990).
\bibitem{Li} J.Q.~Li, Y.~Matsui, S.-K.~Park, and Y.~Tokura, Phys.
  Rev. Lett. {\bf 79}, 297 (1997).
\bibitem{SFO_Bocquet} A.E.~Bocquet, A.~Fujimori, T.~Mizokawa,
  T.~Saitoh, H.~Namatame, S.~Suga, N.~Kimizuka, Y.~Takeda, and
  M.~Takano, Phys. Rev. B {\bf 45}, 1561 (1992).
\bibitem{SrFeO3} T.~Takeda, Y.~Yamaguchi, and H.~Watanabe, J. Phys.
  Soc. Jpn. {\bf 33}, 967 (1972).
\bibitem{Kawasaki} S.~Kawasaki, M.~Takano, R.~Kanno, T.~Takeda, and
  A.~Fujimori, J. Phys. Soc. Jpn., {\bf 67}, 1529 (1998).
\bibitem{Abbate}M.~Abbate, G.~Zampieri, J.~Okamoto, A.~Fujimori,
  S.~Kawasaki, and M.~Takano, Phys. Rev. B{\bf 65}, 165120 (2002).
\bibitem{Takano2} M.~Takano, S.~Nasu, T.~Abe, K.~Yamamoto, S.~Endo,
  Y.~Takeda, and J.B.~Goodenough, Phys. Rev. Lett. {\bf 67}, 3267
  (1991).
\bibitem{Kawakami} T.~Kawakami, S.~Nasu, T.~Sasaki, K.~Kuzushita,
  S.~Morimoto, S.~Endo, T.~Yamada, S.~Kawasaki, and M.~Takano, Phys.
  Rev.  Lett. {\bf 88}, 037602 (2002).
\bibitem{Rozenberg} G.Kh.~Rozenberg, A.P.~Milner, M.P.~Pasternak,
  G.R.~Hearne, and R.D.~Taylor, Phys. Rev. B {\bf 58}, 10283 (1998).
\bibitem{Matsuno} J.~Matsuno, T.~Mizokawa, A.~Fujimori, K.~Mamiya,
  Y.~Takeda, S.~Kawasaki, and M.~Takano, Phys. Rev. B {\bf 60}, 4605
  (1999).
\bibitem{Mizokawa1} T.~Mizokawa and A.~Fujimori, Phys. Rev. B {\bf 54}
  5368 (1996).
\bibitem{Mattheiss} L.F.~Mattheiss, Phys. Rev. B, {\bf 5}, 290
  (1972).
\bibitem{Morimoto} S.~Morimoto, T.~Yamanaka, and M.~Tanaka, Physica B
  {\bf 237-238}, 66 (1997).
\bibitem{Harrison} W.A.~Harrison, {\it Electronic Structure and the
    Properties of Solids} (Dover, New York, 1989)
\bibitem{BKBO_review} S.~Uchida, K.~Kitazawa, and S.~Tanaka, {\it
    Phase Transition} (Gordon and Breach Science, New York, 1987),
  Vol. 8, p.95.
\bibitem{BKBO} A.I.~Liechtenstein, I.I.~Mazin, C.O.~Rodriguez,
  O.~Jepsen, O.K.~Andersen, and M.~Methfessel, Phys. Rev. B {\bf 44},
  5388 (1991).

\end{thebibliography}
\end{document}